\documentclass[twocolumn,showpacs,preprintnumbers,amsmath,amssymb]{revtex4}

\usepackage{graphicx}
\usepackage{dcolumn}
\usepackage{bm}

\begin{document}


\title{Singlet and triplet BCS pairs in a gas of two-species fermionic polar molecules}
\author{T. Shi, J.-N. Zhang, C.-P. Sun, and S. Yi}

\affiliation{Key Laboratory of Frontiers in Theoretical Physics, Institute of Theoretical Physics, Chinese Academy of Sciences, Beijing 100190, China}

\begin{abstract}
We investigate the BCS pairing in a mixture of fermionic polar molecules with two different hyperfine states. We derive a set of coupled gap equations and find that this system supports both spin-singlet and -triplet BCS pairs. We also calculate the critical temperatures and the angular dependence of order parameters. In addition, by tuning short-range interaction between inter-species molecules, the transition between singlet and triplet paired states may be realized.
\end{abstract}

\date{\today}
\pacs{03.75.Ss, 74.20.Rp, 67.30.H-, 05.30.Fk}

\maketitle

{\em Introduction}. --- There have been growing interests in studying the properties of dipolar Fermi gases since the experimental success in making high phase-space-density fermionic KRb gas~\cite{ye1,ye2}. Of particular interest is the anisotropy of electric dipole-dipole interaction (EDDI), whose partially attractive character provides a mechanism to form anisotropic BCS pair. You and Marinescu~\cite{you} first realized that the $p$-wave paired BCS states could be achieved for fermionic atoms inside an external dc field. Baranov {\it et al}. further showed that the order parameter of a BCS pair is the superposition of all odd partial waves~\cite{bara}. Subsequently, the critical temperature of the superfluid transition and its relation to the trap geometry were investigated~\cite{bara}. In those studies, the Fermi surface (FS) of system was assumed to be a sphere. However, it was found via variational method that the Fock exchange interaction deformed the FS~\cite{miya}. This result was also confirmed by both numerical~\cite{ronen,zhang} and perturbation calculations~\cite{ronen,chan}. Very recently, the role played by Fock exchange interaction in BCS pairing is studied~\cite{pu}.

All above mentioned theoretical work focus on the spinless dipolar fermions. Adding the spin degree of freedom, ultracold gas of polar molecules may offer even richer physics. When loaded into an optical lattice, spin-$\frac{1}{2}$ polar molecules can be used as a toolbox for lattice spin models~\cite{zoller}. Furthermore, it was shown that a two-component dipolar Fermi gas may have a ferro-nematic ground state~\cite{freg}.

In the present work, we study the BCS pairing in a mixture of fermionic polar molecules with two different hyperfine states, $\sigma=\uparrow$ and $\downarrow$. The electric dipole moments $d$ of all molecules are orientated along $z$-axis such that the spin-independent EDDI becomes $V_{d}({\mathbf r})=C_dr^{-3}\left(1-\frac{3z^2}{r^2}\right)$ with $C_d=\frac{d^2}{4\pi\varepsilon_0}$. In addition, we assume that inter-species molecules also interact via short-range interaction $V_0({\mathbf r})=C_0\delta({\mathbf r})$, where $C_0=\frac{4\pi\hbar^2a_s}{m}$ with $a_s$ being the $s$-wave scattering length and $m$ the mass of the molecule. The total interaction potential $V=V_d+V_0$ conserves the number of particles in individual spin component. We show that novel {\em spin-singlet even partial wave} and {\em spin-triplet odd partial wave} pairs can appear in this system.

{\em Gap equation}. --- We consider a homogeneous gas of two-species fermionic polar molecules with total number density $2n$. For simplicity, we assume that $n_\uparrow=n_\downarrow=n$. In momentum space, the second quantized Hamiltonian reads
\begin{eqnarray}
H&=&\sum_{{\mathbf k}\sigma}\left(\frac{\hbar^2{\mathbf k}^{2}}{2m}-\mu\right) c_{{\mathbf k}\sigma}^{\dagger}c_{{\mathbf k}\sigma}\nonumber\\
&&+\frac{1}{2{\cal V}} \sum_{{\mathbf k}{\mathbf p}{\mathbf q},\sigma\sigma'} \widetilde V({\mathbf q})c_{{\mathbf k}+{\mathbf q}\,\sigma}^{\dagger }c_{{\mathbf p}-{\mathbf q}\,\sigma'}^{\dagger}c_{{\mathbf p}\sigma'}c_{{\mathbf k}\sigma}, \label{ham}
\end{eqnarray}
where $\mu=\mu_\uparrow=\mu_\downarrow$ is the chemical potential, $\cal V$ is the volume of the system, and $\widetilde V=\widetilde V_d+\widetilde V_0$ is the Fourier transform of the interaction potential. One should keep in mind that, in Eq. (\ref{ham}), contact interaction only exists between $\uparrow$ and $\downarrow$ molecules.

Under Hartree-Fork approximation, Eq. (\ref{ham}) becomes
\begin{eqnarray}
H\!=\!\sum_{{\mathbf k}\sigma}\varepsilon_{{\mathbf k}\sigma} c_{{\mathbf k}\sigma}^{\dagger} c_{{\mathbf k}\sigma}\!+\!\frac{1}{2}\sum_{{\mathbf k},\sigma\sigma'} \Delta_{\sigma\sigma'}(\mathbf k)c_{{\mathbf k}\sigma}^{\dagger} c_{-{\mathbf k}\sigma'}^{\dagger}+h.c.,\label{ham2}
\end{eqnarray}
where the dispersion relation is 
$\varepsilon_{{\mathbf k}\sigma}=\frac{\hbar^2{\mathbf k}^{2}}{2m}-\varepsilon_F-\Sigma_\sigma({\mathbf k})$ with $\varepsilon_F\equiv\mu-nC_0=\frac{\hbar^2}{2M}(6\pi^2n)^{2/3}$ being the Fermi energy and $\Sigma_\sigma({\mathbf k})={\cal V}^{-1}\sum_{\mathbf q}\widetilde V_d({\mathbf k}-{\mathbf q})f_{{\mathbf q}\sigma}$ the self energy. To the lowest order, we choose the Fermi occupation number $f_{{\mathbf q}\sigma}=\left\langle c_{{\mathbf q}\sigma}^\dag c_{{\mathbf q}\sigma}\right\rangle$ to be that at zero temperature. As a result, dispersion relations become spin independent, denoted as $\varepsilon_{\mathbf k}$. The Fock exchange interaction deforms FS such that, in the vicinity of FS, $\varepsilon_{\mathbf k}$ can be approximately expressed as~\cite{miya}
\begin{eqnarray}
\bar\varepsilon_{{\mathbf k}}\simeq\frac{\hbar^2}{2m}\left[\alpha^{-1}(k_{x}^{2}+k_{y}^{2})+\alpha
^{2}k_{z}^{2}\right]-\varepsilon_F,  \label{disp}
\end{eqnarray}
where $\alpha$ is the deformation parameter, which can be found using either variational~\cite{miya} or numerical method~\cite{ronen,zhang}. Moreover, at the weak interaction limit 
\begin{eqnarray}
D\equiv nC_d/\varepsilon_F\ll1,\label{weak}
\end{eqnarray}
the perturbation calculation yields $\alpha=1-\frac{2\pi}{3}D$~\cite{ronen,chan}, where $D$ dimensionless dipole interaction strength. We point out that, for all results presented in this work, $\alpha$ is obtained variationally.

The order parameters for the BCS states in the Hamiltonian (\ref{ham2}) are defined as
\begin{eqnarray}
\Delta_{\sigma\sigma'}({\mathbf k})={\cal V}^{-1}
\sum_{{\mathbf q}}\widetilde V({\mathbf k}-{\mathbf q})\left\langle c_{-{\mathbf q}\sigma'}c_{{\mathbf q}\sigma}\right\rangle.\label{gap}
\end{eqnarray}
Using either Bogoliubov transformation~\cite{he3} or the standard Green's function method~\cite{stat}, we obtain 
\begin{eqnarray}
\left\langle c_{-{\mathbf q}\sigma'}c_{{\mathbf q}\sigma}\right\rangle =-\frac{\Delta_{\sigma\sigma'}({\mathbf q})}{2E_{{\mathbf q}\sigma}}\tanh \left(\frac{1}{2}\beta E_{{\mathbf q}\sigma}\right),\label{mean}
\end{eqnarray}
where $E_{{\mathbf k}\sigma}=\sqrt{\varepsilon_{{\mathbf k}}^{2} +\sum_{\zeta}|\Delta_{\sigma\zeta}({\mathbf k})|^{2}}$ is the energy of single-particle excitation and $\beta=1/(k_BT)$ is the inverse temperature. Substituting Eq. (\ref{mean}) into (\ref{gap}), one find a set of self-consistent equations for order parameters
\begin{eqnarray}
\Delta_{\sigma\sigma'}({\mathbf k})=-\!\int\! \frac{d{\mathbf q}}{(2\pi)^{3}}\widetilde V({\mathbf k}-{\mathbf q})
\frac{\tanh \left(\frac{1}{2}\beta E_{{\mathbf q}\sigma}\right)}{2E_{{\mathbf q}\sigma }}\Delta_{\sigma\sigma'}({\mathbf q}).\label{gap2}
\end{eqnarray}
We note that the order parameters corresponding to different BCS paired states are coupled in above equations. Similar gap equations also appear in superfluid $^3$He system~\cite{he3}. The integral in Eq. (\ref{gap2}) formally diverges. To regularize the interaction, we follow the standard renormalization procedure~\cite{bara}, which, to the first Born approximation, leads to the gap equations
\begin{eqnarray}
\Delta_{\sigma\sigma'}({\mathbf k})&=&-\int \frac{d{\mathbf q}}{(2\pi )^{3}}\widetilde V({\mathbf k}-{\mathbf q})\Delta_{\sigma\sigma'}({\mathbf q})\nonumber\\
&&\quad\times\left[\frac{\tanh\left(\frac{1}{2}\beta E_{{\mathbf q}\sigma}\right)}{2E_{{\mathbf q}\sigma}}-\frac{m}{\hbar^2 {\mathbf q}^{2}}\right].  \label{gap3}
\end{eqnarray}
We emphasize that these equations are only valid in weak interaction regime Eq. (\ref{weak}), as we have ignored the higher order contributions from the interaction.

{\em Symmetries of the order parameters}. --- Up to now, we have expressed the order parameters ${\boldsymbol\Delta}=\left(\begin{array}{cc}
\Delta_{\uparrow\uparrow}&\Delta_{\uparrow\downarrow}\\
\Delta_{\downarrow\uparrow}&\Delta_{\downarrow\downarrow}
\end{array}\right)$ in the uncoupled spin space. While in the basis of coupled spin $\{|SM\rangle\}$, one will have singlet and triplet pairs corresponding to total spin $S=0$ and $1$, respectively. Singlet pairing requires the gap parameter to be antisymmetric such that ${\boldsymbol\Delta}_s=\Delta_si{\boldsymbol\sigma_2}=\left(\begin{array}{cc}
0&\Delta_s\\-\Delta_s&0 \end{array}\right)$, where ${\boldsymbol\sigma}_2$ is the second Pauli matrix. Obviously, $\Delta_s(-{\mathbf k})=\Delta_s({\mathbf k})$, and as we shall show, it is a superposition of even partial waves, i.e., $\Delta_s({\mathbf k})=\sum_{{\rm even}\;l}\Delta_{l}^{(s)}(k)Y_{l0}(\hat{\mathbf k})$.

Following the convention of $^3$He~\cite{he3}, we may define a vector ${\mathbf d}({\mathbf k})$ in spin space by combining three spin components ($M=0,\pm1$) of the triplet pair:
${\boldsymbol\Delta}_t=\sum_{\mu=1,2,3}d_\mu({\boldsymbol\sigma}_\mu i{\boldsymbol\sigma}_2)=\left(\begin{array}{cc}
-d_1+id_2&d_3\\d_3&d_1+id_2 \end{array}\right)$, 
where ${\boldsymbol\sigma}_\mu$ are Pauli matrices. It can be easily seen that
$\Delta_{t,1}=\Delta_{\uparrow\uparrow}=-d_1+id_2$, $\Delta_{t,-1}=\Delta_{\downarrow\downarrow}=d_1+id_2$, and $\Delta_{t,0}=\Delta_{\uparrow\downarrow}=\Delta_{\downarrow\uparrow}=d_3$. In momentum space, $\Delta_{t,M}({\mathbf k})$ are superpositions of odd partial waves, i.e.,  $\Delta_{t,M}({\mathbf k})=\sum_{{\rm odd}\;l}\Delta_{M,l}^{(t)}(k)Y_{l0}(\hat{\mathbf k})$.

{\em Spin-singlet pairing}. --- For deformed FS, we introduce rescaled momentum $\bar{\mathbf k}$ such that $\bar k_{x/y}=\alpha^{-1/2} k_{x/y}$ and $\bar k_{z}=\alpha k_{z}$. Moreover, we define $\xi_{\bar{\mathbf k}}\equiv\hbar^2\bar{\mathbf k}^2/(2M)-\varepsilon_F$, which allows us to rewrite the order parameter as $\Delta_s(\bar{\mathbf k})\equiv\Delta_{s} (\xi_{\bar{\mathbf k}},{\mathbf n}_{\bar{\mathbf k}})$ with ${\mathbf n}_{\bar{\mathbf k}}=\bar{\mathbf k}/\bar{k}$ being an unit vector. Using the fact that pairing is mainly contributed by states near FS, we introduce a characteristic energy $\overline\omega_s$ (of the order of $\varepsilon_F$) to single out the contribution from $-\overline\omega_s\leq\xi_{\bar{\mathbf k}}\leq\overline\omega_s$~\cite{bara}. In weak coupling regime, we have $\overline\omega_s\gg|\Delta_s|,\,k_BT_c^{(s)}$. The value of $\overline\omega_s$ will be determined self-consistently. After some tedious calculations, we obtain from Eq.~(\ref{gap3}) that
\begin{eqnarray}
&&\!\!\!\!\!\!\!\!\!\!\!\!\Delta_{s}(\xi_{\bar{\mathbf k}},{\mathbf n}_{\bar{\mathbf k}})=-\int \frac{d{\mathbf n}_{\bar{\mathbf q}}}{4\pi}W(T,{\mathbf n}_{\bar{\mathbf q}
}){\cal R}_s(\xi_{\bar{\mathbf k}},{\mathbf n}_{\bar{\mathbf k}};0,{\mathbf n}_{\bar{\mathbf q}})\nonumber\\
&&\!\!\!\!\times\Delta_{s}(0,{\mathbf n}_{\bar{\mathbf q}})+\frac{1}{2}\int \frac{d{\mathbf n}_{\bar{\mathbf q}}}{4\pi}\int_{-\varepsilon_F}^{\infty }d\xi_{\bar{\mathbf q}}\ln\left(\frac{\left|\xi_{\bar{\mathbf q}}\right| +\eta_{\bar{\mathbf q}}}{2\overline{\omega}_s}\right)\nonumber\\
&&\!\!\!\!\times\frac{d}{d|\xi_{\bar{\mathbf q}}|}{\cal R}_s(\xi_{\bar{\mathbf k}},{\mathbf n}_{\bar{\mathbf k}}; \xi_{\bar{\mathbf q}},{\mathbf n}_{\bar{\mathbf q}})\Delta_{s}(\xi_{\bar{\mathbf q}},{\mathbf n}_{\bar{\mathbf q}})\nonumber\\
&&\!\!\!\!-\frac{1}{2}\int\!\! \frac{d{\mathbf n}_{\bar{\mathbf q}}}{4\pi }\ln \frac{
|\varepsilon_F|}{\overline\omega_s}{\cal R}(\xi_{\bar{\mathbf k}},{\mathbf n}_{\bar{\mathbf k}};-\varepsilon_F,{\mathbf n}_{\bar{\mathbf q}})\Delta_s(-\varepsilon_F,{\mathbf n}_{\bar{\mathbf q}}),
\label{gaptc}
\end{eqnarray}
where $W(T,{\mathbf n}_{\bar{\mathbf q}})=\int_{0}^{\overline\omega_s} d\xi_{\bar{\mathbf q}}\eta_{\bar{\mathbf q}}^{-1}\tanh[\eta_{\bar{\mathbf q}\sigma}/(2k_BT)]$ and $\eta_{\bar{\mathbf k}}=\sqrt{\xi_{\bar{\mathbf k}}^{2}+|
\Delta_{s}(0,{\mathbf n}_{\bar{\mathbf k}})|^{2}}$. Furthermore, the integration kernel takes the form
\begin{eqnarray}
&&{\cal R}_s\left(\xi_{\bar{\mathbf k}},{\mathbf n}_{\bar{\mathbf k}}; \xi_{\bar{\mathbf q}},{\mathbf n}_{\bar{\mathbf q}}\right)=\frac{
m\bar{\mathbf q}}{2\pi ^{2}} \widetilde V({\mathbf k}-{\mathbf q})\nonumber\\
&&\qquad\times\left[ \frac{\eta_{\bar{\mathbf q}}}{E_{{\mathbf q}}}\frac{\tanh\left(\frac{1}{2}\beta E_{{\mathbf q}}\right)}{\tanh \left(\frac{1}{2}\beta \eta_{\bar{\mathbf q}}\right)}-\frac{
2M\eta_{\bar{\mathbf q}}}{{\mathbf q}^{2}\tanh \left(\frac{1}{2}\beta \eta_{\bar{\mathbf q}}\right)}\right].\nonumber
\end{eqnarray}

Close to the critical temperature $T_{c}^{(s)}$, the gap goes to zero and $\eta_{
\bar{\mathbf q}}\rightarrow |\xi_{\bar{\mathbf q}}|$. For the first term in rhs of Eq. (\ref{gaptc}), we have ${\cal C}\equiv W\left(T_c^{(s)},{\mathbf n}_{\bar{\mathbf k}}\right)=\ln \left[2e^{\gamma}\overline\omega_s/\left(\pi k_BT_{c}^{(s)}\right)\right]$ with $\gamma=0.5772$ being the Euler constant. In contrast, the last two terms in Eq.~(\ref{gaptc}) do not contain the large logarithm $\ln \left(\overline\omega_s/k_BT_{c}^{(s)}\right)$, indicating that they are only important for the preexponential factor in the expression of critical temperature~\cite{bara}. For the purpose of determining $T_c^{(s)}$, they are neglected such that we find on FS
\begin{eqnarray}
\Delta_{s}(0,{\mathbf n}_{\bar{\mathbf k}})=-{\cal C}\int \frac{d{\mathbf n}_{\bar{\mathbf q}}}{4\pi}{\cal R}_s(0,{\mathbf n}_{\bar{\mathbf k}};0,{\mathbf n}_{\bar{\mathbf q}})\Delta_{s}(0,{\mathbf n}_{\bar{\mathbf q}}), \label{eigen}
\end{eqnarray}
which is essentially an eigenvalue equation of integral operator with kernel ${\cal R}_s(0,{\mathbf n}_{\bar{\mathbf k}};0,{\mathbf n}_{\bar{\mathbf q}})$. Finding the highest critical temperature is equivalent to finding the lowest negative eigenvalue of this eigenvalue equation. Even in case $\alpha\neq 1$, it can be shown that the eigenstates corresponding to lowest eigenvalue is independent of the azimuthal angle of ${\mathbf n}_{\bar{\mathbf k}}$~\cite{bara}. This observation allows us to integrate over the azimuthal variable of Eq. (\ref{eigen}) to obtain an eigenvalue equation
\begin{equation}
\int_{-1}^{1}dy{\cal K}_{s}(x,y)\phi(y)=\lambda\phi(x),\label{gaptc2}
\end{equation}
where $x=\cos\theta_{{\mathbf n}_{\bar{\mathbf k}}}$, $y=\cos\theta_{{\mathbf n}_{\bar{\mathbf q}}}$, and ${\cal K}_{s}(x,y)=\big[3l_\alpha(x,y)\,|x-y|-1\big] +k_{F}a_s/(\pi^2D)$ with $l_\alpha\equiv\left\{4\alpha ^{3}+(1-\alpha ^{3})[(x-y)^{2}-\alpha ^{3}(x+y)^{2}]\right\}^{-1/2}$ and $k_F=\sqrt{2M\varepsilon_F}/\hbar$ being the Fermi wave vector. Clearly, the first term in ${\cal K}_s$ originates from EDDI, while the second one corresponds to contact interaction. 

\begin{figure}
\centering
\includegraphics[width=2.4in]{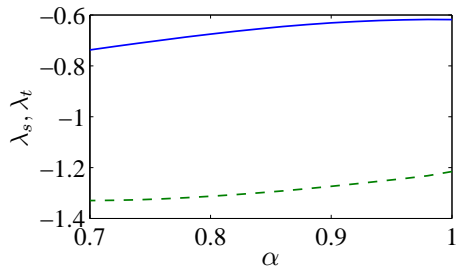}
\caption{The $\alpha$ dependence of $\lambda_s$ for $k_Fa_s=0$ (solid line) and $\lambda_t$ (dashed line).} \label{lmd}
\end{figure}

For singlet pairing, we focus on the subspace of the eigenstates with even parity, i.e., $\phi(x)=\phi(-x)$. Assuming that $\lambda_s$ is lowest eigenvalue of Eq. (\ref{gaptc2}) corresponding to an even eigenstate $\phi_s(x)$, the critical temperature can then be expressed as
\begin{equation}
k_BT_{c}^{(s)}=\frac{2e^{\gamma}\overline\omega_{s}}{\pi}\exp \left(-
\frac{1}{\pi D |\lambda_{s}|}\right).
\end{equation}
In addition, at $T_c^{(s)}$, the order parameter on FS becomes
$\Delta_s(0,{\mathbf n}_{\bar{\mathbf k}})=\Delta_{s0}\phi_s\left(\cos\theta_{{\mathbf n}_{\bar{\mathbf k}}}\right)$, where $\phi_s(\cos\theta_{{\mathbf n}_{\bar{\mathbf k}}})$ represents the angular dependence of the order parameter on FS.

If the Fock exchange interaction is ignored, we have exactly $\phi_{s}(x)=\mathcal{N}\cos\left(x\sqrt{3/|\lambda_{s}|}\right)$, where $\mathcal{N}$ is a normalization constant and $|\lambda_{s}|$ is the largest positive root of the equation $t\cos\sqrt{3/t}+(1+\frac{2k_{F}a_s}{\pi ^{2}D })\sqrt{t/3}\sin\sqrt{3/t}=0$. It can be shown that $\lambda_s$ ($<0$) is a decreasing function of $k_Fa_s/D$. For negative $k_Fa_s$, our result reduces to that for $s$-wave pairing when $D=0$. In case $\alpha\neq1$, $\lambda_s$ and $\phi_s$ can only be determined numerically. As shown in Fig. \ref{lmd}, $\lambda_s$ is an increasing of $\alpha$ for $k_Fa_s=0$, indicating that, in the absence of contact interaction, the Fock exchange interaction enhances singlet BCS pairing.

\begin{figure}
\centering
\includegraphics[width=3.4in]{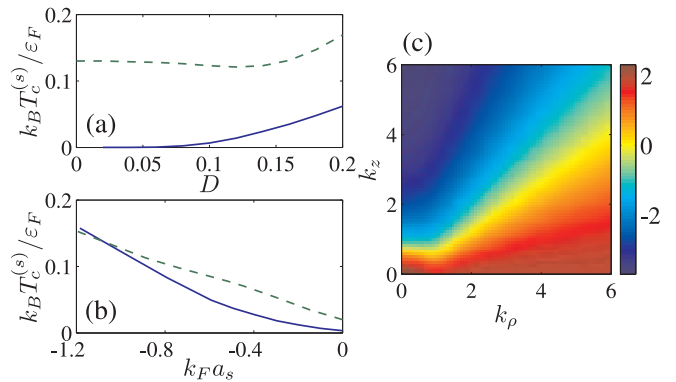}
\caption{(color online) (a) The $D$ dependence of $T_c^{(s)}$ for $k_Fa_s=0$ (solid line) and $-1$ (dashed line). (b) $T_c^{(s)}$ as a function of $k_Fa_s$ for $D=0.1$ (solid line) and $0.15$ (dashed line). (c) $\Delta_s(k_\rho,k_z)/\Delta_{s0}$ at critical temperature for $k_Fa_s=0$ and $D=0.01$, where $k_\rho=(k_x^2+k_y^2)^{1/2}$.} \label{slt}
\end{figure}

Next, we compute the characteristic energy $\overline\omega_s$. Following the Gor'kov and Melik-Barkhudarov's approach~\cite{bara,gm}, we obtain
\begin{eqnarray}
&&\!\!\!\!\!\!\!\!\ln \frac{\overline{\omega}_s}{\varepsilon_F}=\ln 0^{+}+\frac{1}{2\pi D \lambda_{\mathrm{s}}}\int \frac{d{\mathbf n}_{\bar{\mathbf k}}}{
4\pi }\int \frac{d{\mathbf n}_{\bar{\mathbf q}}}{4\pi }\phi_{s}({\mathbf n}_{\bar{\mathbf k}}) 
\nonumber\\
&&\quad\times \int_{-\varepsilon_F}^{\infty }\frac{d\xi_{\bar{\mathbf q}}}{|\xi_{\bar{\mathbf q}}|}{\cal R}_{s}(0,{\mathbf n}_{\bar{\mathbf k}};\xi_{\bar{\mathbf q}},{\mathbf n}_{
\bar{\mathbf q}})\frac{\Delta_{s}(\xi_{\bar{\mathbf q}},{\mathbf n}_{\bar{\mathbf q}})}{\Delta_{0}},\label{omgb}
\end{eqnarray}
where the order parameter, at $T_c^{(s)}$, satisfies the equation
\begin{eqnarray}
\frac{\Delta_{s}(\xi_{\bar{\mathbf k}},{\mathbf n}_{\bar{\mathbf k}})}{\Delta _{s0}}\!\simeq\!
\frac{1}{\pi D \lambda_{s}}\!\int\!\! \frac{d{\mathbf n}_{\bar{\mathbf q}}}{4\pi }
{\cal R}_s(\xi_{\bar{\mathbf k}},{\mathbf n}_{\bar{\mathbf k}};0,{\mathbf n}_{\bar{\mathbf q}})\phi_{s}(\cos\theta_{\bar{\mathbf q}}).\label{order}
\end{eqnarray}
$\overline\omega_s$ can be obtained by substituting Eq. (\ref{order}) into (\ref{omgb}). We remark that, in the rhs of Eq. (\ref{omgb}), the second term also contains a divergent term which cancels the divergence from the first term. In general, $\overline\omega_s$ is a function of both $D$ and $k_Fa_s$, indicating that it also implicitly depends on $\alpha$. For simplicity, we use $\alpha=1$ to calculate $\overline\omega_s$, which yields $\overline\omega_s\simeq 0.36\varepsilon_F$ for $k_Fa_s=0$ and $\overline\omega_s\simeq 0.54\varepsilon_F$ for $D=0$. The latter is exactly the value for $s$-wave pairing.

In Fig.~\ref{slt} (a), we present the $D$ dependence of the critical temperature for singlet pairing. For $k_Fa_s=0$, only dipolar interaction is responsible for pairing, therefore, $T_c^{(s)}$ is a monotonically increasing function of $D$. However, for $k_Fa_s=-1$, $T_c^{(s)}(D)$ becomes a concave curve. This result suggests that even though both attractive contact interaction and dipolar interaction contribute to pairing, they also compete with each other. In fact, the deformed FS makes it difficult for contact interaction to form $s$-wave pairs, which is responsible for the drop of $T_c^{(s)}$ at small $D$. When dipolar interaction dominates, $T_c^{(s)}$ becomes a increasing function of $D$ again. We also verify that if the FS deformation is ignored,  $T_c^{(s)}$ would always be increasing function of $D$. Figure~\ref{slt} (b) shows the $k_Fa_s$ dependence of $T_c^{(s)}$ for given $D$. As $\alpha$ is fixed by $D$, $T_c^{(s)}$ always increases as one increases the strength of contact interaction. In addition, when short-range interaction dominates, the one with smaller FS deformation (smaller $D$) will eventually has a higher critical temperature. Finally, the angular dependence of order parameter for spin-singlet pair at critical temperature is presented in Fig.~\ref{slt}~(c).

{\em Spin-triplet pairing}. --- At critical temperature, the order parameters for triplet pairs $\Delta_{t,M}$ are decoupled. In addition, they satisfy gap equations with exactly the same form. Therefore, we denote the triplet order parameter as $\Delta_t$. Since the critical temperature and order parameters can be obtained by following the same procedure as that for singlet pairing, here we shall only present our results.

\begin{figure}
\centering
\includegraphics[width=3.4in]{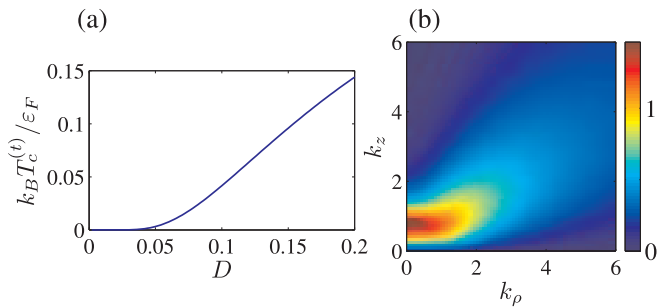}
\caption{(color online) (a) The $D$ dependence of $T_c^{(t)}$. (b)  $\Delta_t(k_\rho,k_z)/\Delta_{t0}$ at critical temperature for $D=0.01$} \label{tlt}
\end{figure}

The critical temperature $T_c^{(t)}$ for triplet pairing can be expressed as $k_BT_{c}^{(t)}=\frac{2e^{\gamma}\overline\omega_{t}}{\pi }\exp\left(-\frac{1}{\pi D |\lambda_{t}|}\right)$, where $\lambda_t$ is the lowest eigenvalue of the equation $\int_{-1}^{1}dy{\cal K}_t(x,y)\phi (y)=\lambda\phi (x)$ in the subspace of the eigenstates with odd parity. The integration kernel is $
{\cal K}_{t}(x,y)=3l_\alpha(x,y)\,|x-y|-1$, which is essentially ${\cal K}_s$ with $k_Fa_s=0$. Assuming that the corresponding eigenstate of $\lambda_t$ is $\phi_t(x)$, the triplet order parameter on FS is
$\Delta_t(0,\cos\theta_{{\mathbf n}_{\bar{\mathbf k}}})=\Delta_{t0}\phi_t(\cos\theta_{{\mathbf n}_{\bar{\mathbf k}}})$. The $\alpha$ dependence of $\lambda_t$ (Fig.~\ref{lmd}) indicates that the triplet pairing is also enhanced by the Fock exchange interaction. 

\begin{figure}
\centering
\includegraphics[width=2.4in]{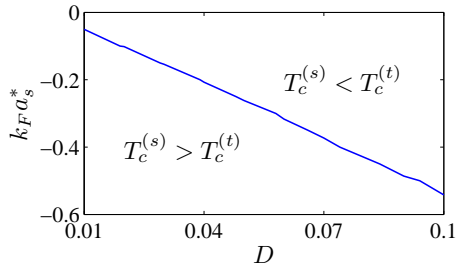}
\caption{The dipolar interaction strength dependence of the critical scattering length.} \label{phase}
\end{figure}

The characteristic energy $\overline\omega_t$ and the triplet order parameter $\Delta_t(\xi_{\bar{\mathbf k}},{\mathbf n}_{\bar{\mathbf k}})/\Delta_{t0}$ at critical temperature satisfy similar equations as Eqs. (\ref{omgb}) and (\ref{order}), except for that ${\cal R}_s$ is now replaced by ${\cal R}_t$ which does not contain the contact interaction. In Fig. \ref{tlt} (a), we plot the dipolar interaction strength dependence of the critical temperature $T_c^{(t)}$, where we have used $\overline\omega_t\simeq0.42\varepsilon_F$, the value corresponding to $\alpha=1$ case. The typical order parameter for triplet pair is plotted in Fig.~\ref{tlt} (b). We point out that in case the FS deformation is ignored, our results for triplet pairing reduce to those obtained by Baranov {\it et al}.~\cite{bara}.

{\em Singlet versus triplet pairing}. --- For $k_Fa_s=0$, it can be seen from Fig. \ref{lmd} that we always $|\lambda_s|<|\lambda_t|$; by further assuming $\alpha=1$, we also find that $\overline\omega_s<\overline\omega_t$. These results suggest that, without attractive contact interaction, the critical temperature for singlet pairing is always lower than that for triplet pairing. However, for $k_Fa_s<0$, $T_c^{(s)}$ increases as one increases the strength of short-range interaction such that beyond a threshold $k_Fa_s^*$, we have $T_c^{(s)}>T_c^{(t)}$. The critical $s$-wave scattering length is determined by equation $\pi D\ln \frac{\overline\omega_s(k_Fa_s^*)}{\overline\omega_t}= \frac{1}{\lambda_t}-\frac{1}{\lambda_s(k_Fa_s^*)}$, where we have explicitly expressed $\overline\omega_s$ and $\lambda_s$ as functions of $k_Fa_s$. In Fig. \ref{phase}, we present the $D$ dependence of the critical scattering length. We point out that when $T_c^{(s)}=T_c^{(t)}$, the singlet and triplet pairs may coexist in the system.

{\em Conclusion}. --- We have studied the BCS pairing in a ultracold gas of two-species fermionic polar molecules. We show that there exist two types of BCS pairs: spin-singlet even partial wave and spin-triplet odd partial pairs. We calculate the critical temperatures and angular dependence of the order parameters. It is found that, without attractive contact interaction, the critical temperature for triplet pairing is always higher than that for singlet pairing. The opposite could also be true if one tunes the scattering length to a large negative value. Compared to the spin-triplet $p$-wave pairing in superfluid $^3$He, the spin-triplet (-singlet) pairing studied here couples all odd (even) partial waves. In addition, by carefully tuning the contact interaction, singlet and triplet pairs may coexist in this system. Armed with the rich experimental control mechanisms developed in cold atom physics, two-component fermionic polar molecules may provide an ideal platform for exploring new physics beyond what superfluid $^3$He could offer. 

This work was supported by NSFC through grants 10974209 and 10935010 and by the National 973 program (Grant No. 2006CB921205). S.Y. acknowledges the support of the ``Bairen" program of Chinese Academy of Sciences.


\begin{thebibliography}{}
\bibitem{ye1} S. Ospelkaus {\it et al}., 
Nat. Phys. {\bf4}, 622 (2008).

\bibitem{ye2} K.-K. Ni {\it et al}., 
Science {\bf322}, 231 (2008).

\bibitem{you} L. You and M. Marinescu, Phys. Rev. A {\bf 60}, 2324 (1999).

\bibitem{bara} M. A. Baranov {\it et al}., 
Phys. Rev. A {\bf66}, 013606 (2002); M.A. Baranov {\it et al}., 
Phys. Rev. Lett. {\bf92}, 250403 (2004); M. A. Baranov {\it et al}., 
Phys. Rev. Lett. {\bf94}, 070404 (2005).

\bibitem{miya} T. Miyakawa {\it et al}., 
Phys. Rev. A {\bf77}, 061603 (2008).

\bibitem{ronen} S. Ronen and J. L. Bohn, arXiv:0906.3753 (2009).

\bibitem{zhang} J.-N. Zhang and S. Yi, arXiv:0907.2804 (2009).

\bibitem{chan} C.-K. Chan {\it et al}., 
arXiv:0906.4403 (2009).

\bibitem{pu} C. Zhao {\it et al}., unpublished.

\bibitem{zoller} A. Micheli {\it et al}., 
Nat. Phys. {\bf 2}, 341 (2006).

\bibitem{freg} B. M. Fregoso and E. Fradkin, arXiv:0907.1345 (2009).

\bibitem{he3} D. Vollhardt and P. W\"{o}lfle, {\em The superfluid phases of
Helium 3} (Taylor \& Francis, London, 1990).

\bibitem{stat} See e.g. E. M. Lifshitz and L. P. Pitaevskii, {\em Statistical Physics Part 2: Theory of the Condensed State} (Pergamon Press, Oxford, 1980).

\bibitem{gm} L. P. Gor'kov and T.K. Melik-Barkhudarov, Zh. \'{E}ksp. Teor.
Fiz. 40, 1452 (1961) [Sov. Phys. JETP 13, 1018 (1961)].

\end{thebibliography}
\end{document}